\documentstyle[prl,eqsecnum,aps]{revtex}

\begin{document}
\author{Jian-Qi Shen \footnote{E-mail address: jqshen@coer.zju.edu.cn}}
\address{ Center for Optical and Electromagnetic Research, State Key Laboratory of Modern Optical
Instrumentation, College of Information Science and Engineering;
Zhejiang Institute of Modern Physics and Department of Physics,
Zhejiang University, Hangzhou 310027, P. R. China}
\date{\today }
\title{Anti-shielding Effect and Negative Temperature in Instantaneously Reversed Electric Fields and Left-Handed Media}
\maketitle

\begin{abstract}
The connections between the anti-shielding effect, negative
absolute temperature and superluminal light propagation in both
the instantaneously reversed electric field and the left-handed
media are considered in the present paper. The instantaneous
inversion of the exterior electric field may cause the electric
dipoles into the state of negative absolute temperature and
therefore give rise to a negative effective mass term of
electromagnetic field ( i. e., the electromagnetic field
propagating inside the negative-temperature medium will acquire an
imaginary rest mass ), which is said to result in the potential
superluminality effect of light propagation in this anti-shielding
dielectric. In left-handed media, such phenomena may also arise.

 PACS Ref: 78.20.Nv; 05.20.-y; 78.20.Ci

\end{abstract}


\section{Introduction}
This paper is concerned with the physically interesting
anti-shielding effects of electric and magnetic dipoles, and the
related phenomena such as the negative absolute temperature of
dipole systems and the consequent superluminal light propagation.
In nature, neither of the above three effects ( phenomena ) is
easily observed. Although negative magnetic permeability has been
shown to be possible when a polariton resonance exists in the
permeability, such as in the antiferromagnets MnF$_{2}$ and
FeF$_{2}$\cite{Camley}, or certain insulating
ferromagnets\cite{Hartstein}, a negative permeability ($\mu$) with
low losses coexisting with a negative electric permittivity
($\epsilon$) has not been demonstrated\cite{Smith}. This means
that in these magnetic anti-shielding materials ( $\mu <0,\epsilon
>0$ ), electromagnetic wave cannot propagate through it, since the
negative magnetic permeability and positive electric permittivity
may result in an imaginary index of refraction. The principal
reason for the state of negative absolute temperature being rarely
observed in nature lies in that the attainment of negative
temperature should be restricted to some critical
conditions\cite{Ramsey} such as that the system must possess an
upper limit to the possible energy level of the allowed states;
and the system should be thermally isolated from other system at
positive temperature, for if a negative- and positive- temperature
systems are in thermal contact, energy will be given up from the
negative temperature to the positive, and therefore the
negative-temperature system will decay to a normal state (
positive temperature ) through infinite temperature. In history,
only the experiments of nuclear spin systems of some crystals,
where the state of negative absolute temperature might be
attained, were carried out\cite{Purcell,Klein}.

What is referenced above is in connection only with the
anti-shielding effect and negative temperature of magnetic media;
the light propagation in electric and magnetic anti-shielding
media is not involved in these investigations. Here, however, we
take into account the anti-shielding effect of electric media and
the light propagation inside it, i. e., we deal with the
statistical mechanics of the state of negative temperature caused
by the instantaneous reversal of the exterior electric field and,
based on this, briefly discuss the {\it negative effective rest
mass term} of light due to negative temperature and the consequent
superluminality effect in the anti-shielding dielectric. Here the
negative effective rest mass term of light means that the
electromagnetic field propagating inside the anti-shielding medium
would acquire an imaginary rest mass. Although it sounds somewhat
surprising that in the anti-shielding dielectric there exists the
superluminal light propagation, this superluminality motion
associated with the particle velocity rather than with the group
velocity of the light field is actually familiar to us, since it
is in connection with the  `` tachyon '' representation in the
Poincar$\acute{\rm e}$ group\cite{Ramond} and is therefore said to
be theoretically natural.

In the present paper, the potential fact that the negative
absolute temperature of dipole systems of left-handed media is
also demonstrated. Left-handed media is such a hypothetical
material which has a simultaneously {\it negative electric
permittivity } and {\it magnetic permeability}. In 1968,
Veselago\cite{Veselago} predicted that electromagnetic plane wave
in a left-handed medium would propagate in a direction opposite to
that of the flow of energy. It is of interest that the refractive
index, $n$, of this medium is also negative, which can be
rigorously derived by analyzing the boundary conditions of
incident waves on the interface between left-handed and ordinary
media. This, therefore, means that on the interface between air
and left-handed media, anomalous refraction takes
place\cite{Veselago,Klimov}. It is apparently seen that in this
media where both permittivity and permeability are taken to be
negative, the anti-shielding effect arises.

We hold that there exists close relationship between the
electromagnetic anti-shielding effect and the negative absolute
temperature as well as the potential superluminal light
propagation due to anti-shielding effects. The superluminal light
propagation presented in this paper results from the statistical
fluctuations of polarization charge in negative-temperature media,
as will be discussed in Sec. 3. Since in the present paper we
concern ourselves much with the light propagation in media, in
next section we propose a phenomenological theory of {\it
effective rest mass term} of light field, which is helpful in
considering the light propagation in anti-shielding and
negative-temperature media. The viewpoint of {\it effective rest
mass} of electromagnetic field in media is useful in giving the
vivid descriptions of some physical effects, for example, the
Meissner effect in superconductivity can be interpreted as
follows: the superconducting electrons, which produce the
self-induced charge current\cite{Hou}, can provide the
electromagnetic field with an effective rest mass ($m_{eff}$), and
this nonvanishing $m_{eff}$ is therefore responsible for the fact
that the magnetic field penetrates only a short distance (
proportional to $m_{eff}^{-1}$ ) within the surface of a
superconductor. The interaction of light wave with media is one of
the leading subjects in classical electrodynamics and quantum
optics, where a multitude of physical processes can often be
phenomenologically described by useful models at classical and/or
quantum level\cite{Shen1,Shen2,Shen3}.

\section{Effective rest mass term of light field}
In history, possibility of a light pulse with speed greater than
that ($c$) in a vacuum has been extensively investigated by many
authors\cite {Akulshin,Bolda,Mitchell,Ziolkowski,Zhou,Nimtz1}. The
various experiments with microwave signals performed by Nimtz {\it
et al.} revealed superluminal velocities of evanescent modes\cite
{Nimtz2,Nimtz3,Nimtz4}. Wang {\it et al.} used the method of
gain-assisted linear anomalous dispersion to demonstrate the
superluminal light propagation in the atomic caesium gas and
measured a negative group-velocity index of the medium that means
the faster-than-$c$ propagation of light\cite {Wang,Marangos}. We
suggest that, based on the theory of effective rest mass of light
field, Wang's superluminal propagation due to the anomalous
dispersion in the gaseous atomic medium\cite {Wang} can be
phenomenologically understood as follows:

The relation between the frequency $\omega $ and wave vector $k$
of lightwave propagating inside the dispersive medium may be
$\frac{\omega ^{2}}{k^{2}}=\frac{c^{2}}{n^{2}}$ with $n$ being the
optical refractive index. For highly absorptive media, the
corresponding modification of this method can also apply. So, for
convenience, here we are not ready to deal with the case of
absorptive materials. This dispersion relation $\frac{\omega
^{2}}{k^{2}}=\frac{c^{2}}{n^{2}}$ yields

\begin{equation}
\frac{\omega ^{2}}{k^{2}}=\frac{c^{2}}{1-\frac{c^{4}}{\hbar
^{2}}\frac{m^{2}}{\omega ^{2}}}               \label{eq101}
\end{equation}
with the effective rest mass, $m$, of the light being so defined
that $m$, $n$ and $\omega $ together satisfy the following
relation

\begin{equation}
\frac{m^{2}c^{4}}{\hbar ^{2}}=(1-n^{2})\omega ^{2}.
                 \label{eq102}
\end{equation}
This, therefore, implies that the particle velocity of light (
photons ) is $v=nc$ and the phase velocity of light is
$v_{p}=\frac{c}{n}=\frac{c^{2}}{v}$. Since this relation is
familiar to us, we thus show that the above calculation is
self-consistent, at least for the de Broglie wave ( particle ).

As an illustrative example of the application of the above
formulation, we now evaluate the effective rest mass of lightwave
propagating in some certain media. For instance, we consider the
case of electron plasma in which the relative permittivity
$\epsilon (\omega )$ reads\cite{Cairns}

\begin{equation}
\epsilon (\omega )=1-\frac{\omega _{p}^{2}}{\omega ^{2}}, \quad
 \omega _{p}^{2}=\frac{N(e^{\ast
})^{2}}{m\epsilon _{0}}  \label{eqq103}
\end{equation}
with $\omega _{p},N,e^{\ast },m$ and $\epsilon _{0}$ being the
electron plasma frequency, number of electrons per unit volume,
effective charge of electrons, electron mass and electric
permittivity in a vacuum, respectively. Note that here the
electron plasma is regarded as cold plasma where the thermal
effects can be ignored. Since the relative magnetic permeability,
$\mu $, of this plasma can be approximately taken to be $1$ and in
consequence the refractive index $n(\omega )=\sqrt{\epsilon
(\omega )}$, it follows from (\ref {eq102}) and (\ref {eqq103})
that the effective rest mass of the incident electromagnetic wave
( $\omega
>\omega _{p}$ ) is of the form

\begin{equation}
m_{eff}=\frac{\hbar \omega _{p}}{c^{2}},
\end{equation}
which is independent of the frequency of the electromagnetic wave
propagating inside this plasma. It is of interest that the
effective $\epsilon (\omega )$ of the array of long metallic wires
(ALMWs)\cite{Pendry1,Pendry2,Pendry3,Maslovski} ( that is now used
to manufacture the left-handed materials ) behaves as that of
plasma, which means that in ALMWs media ( $\epsilon (\omega
)=1-\frac{\omega _{0}^{2}}{\omega ^{2}},\mu =1$ ), the lightwave
may also acquire an $\omega $- independent effective rest mass.
However, in most dielectric, e. g., the anomalous dispersive
media\cite{Akulshin} where the permittivity $\epsilon (\omega )$
cannot be taken to be the plasma type, the propagating lightwave
may therefore acquire an $\omega $- dependent effective rest mass.
In what follows we calculate the group velocity in order to
consider the superluminal light propagation in anomalous
dispersive media.

It follows that the group-velocity index is

\begin{equation}
n_{g}=\frac{1}{n}-\frac{mc^{3}}{\hbar ^{2}k}\frac{{\rm d}m}{{\rm
d}\omega }, \label{eeq25}
\end{equation}
which can yield the familiar formula

\begin{equation}
n_{g}=n+\omega \frac{{\rm d}n}{{\rm d}\omega },
\end{equation}
so long as we substitute the expression for the effective rest
mass $m$ into the expression (\ref{eeq25}). Thus, the group
velocity of light pulse is

\begin{equation}
v_{g}=\frac{c}{n_{g}}=\frac{nc}{1-\frac{mnc^{3}}{\hbar
^{2}k}\frac{{\rm d}m}{{\rm d}\omega }}.
\end{equation}
This, therefore, means that if the rest mass $m$ is independent of
$\omega$, then the particle velocity $v=nc$ equals the group
velocity $v_{g}$, which is analogous to the case of the de Broglie
particle; nevertheless, if the effective rest mass $m$ of the
light in the dispersive medium depends strongly on $\omega $ ( e.
g., the property caused in the anomalous-dispersive medium ), then
the group velocity $v_{g}$ is completely divorced from the
particle velocity $v$ and may be greater than $c$ or even
negative. Here both the particle velocity and the group velocity
refer to those of the same matter wave that possesses the
wave-particle duality. Note, however, that although the group
velocity $v_{g}$ of light propagation exceeds $c$ of the speed of
light in a vacuum, this superluminal phenomenon is not related
closely to the special relativity since in this case the particle
velocity of photons does not often necessarily exceed $c$. In the
present paper, from the point of view of effective rest mass we
suggest a new possible superluminality effect of light in the
negative-temperature medium ( i. e., the dielectric placed in the
instantaneously reversed electric fields ). Since it is connection
with the particle velocity, rather than with the group velocity
measured in Wang's superluminal experiment, this superluminal
light propagation in the present paper is related closely to the
dynamics of special relativity, in other words, the particle
velocity ( rather than the group velocity ) of light ( photons )
propagating through the state of negative temperature may exceed
$c$ of the photon in a vacuum; the negative effective rest mass
term of light results from the state of negative absolute
temperature that is caused by the electric anti-shielding effect (
due to the instantaneous reversals of external electric fields ).

\section{Anti-shielding effect of electric dipoles in state of negative temperature}
The potential energy of the electric dipole moment $p_{0}$ in the
exterior electric field $E$ is $\varepsilon _{i}=\frac{p_{0}D\cos
\theta _{i}}{\epsilon \epsilon _{0}}$, where $\epsilon$ and
$\epsilon _{0}$ are respectively the relative permittivity of
dielectric and the permittivity of free space; $D$ stands for the
electric displacement vector. $\theta _{i}$ is taken to be in the
region $0\leq \theta _{i}<\frac{\pi }{2}$. We regard the
continuous angle $\theta $ between dipole moment and exterior
field as the discrete $\theta _{i}$ in order to simplify the
following calculations. This means the index $i$ ranges from $1$
to $\infty $. Let $N_{i}^{+}$ and $N_{i}^{-}$ respectively denote
the occupation numbers of the electric dipoles with the potential
energy $\varepsilon _{i}$ and $-\varepsilon _{i}$. Thus the total
number $N$ of electric dipoles in the dielectric medium and the
total potential energy $U$ of these dipoles in exterior field are
given as follows

\begin{equation}
N=\sum_{i}\left ({N_{i}^{+}+N_{i}^{-}}\right ), \quad
U=\sum_{i}\left({N_{i}^{+}-N_{i}^{-}}\right )\varepsilon _{i}.
\label{eq17}
\end{equation}
$N_{i}^{+}$ and $N_{i}^{-}$ are solved in what follows. For
convenience, we set $N_{i}^{+}=\frac{1}{2}(a_{i}+b_{i})$ and $
N_{i}^{-}=\frac{1}{2}(a_{i}-b_{i})$, and then we have
$a_{i}=N_{i}^{+}+N_{i}^{-}, \quad b_{i}=N_{i}^{+}-N_{i}^{-}$ and
$N=\sum_{i}a_{i},\quad U=\sum_{i}b_{i}\varepsilon _{i}$. It is
apparent that $N$ can be rewritten

\begin{equation}
N=N\frac{\sum_{i}c_{i}}{\sum_{j}c_{j}}+\sum_{i}f_{i}
\end{equation}
with $\sum_{i}f_{i}=0$. Since $N=\sum_{i}a_{i}$, it follows that
$a_{i}=N\frac{c_{i}}{\sum_{j}c_{j}}+f_{i}$.

In the similar fashion, $U$ can be rewritten

\begin{equation}
U=\frac{\sum_{i}d_{i}}{\sum_{j}d_{j}}U+\sum_{i}g_{i}\varepsilon
_{i}
\end{equation}
with $\sum_{i}g_{i}\varepsilon _{i}=0$. Since
$U=\sum_{i}b_{i}\varepsilon _{i}$, it follows that $
b_{i}\varepsilon
_{i}=\frac{d_{i}}{\sum_{j}d_{j}}U+g_{i}\varepsilon _{i}$ or $
b_{i}=\frac{1}{\sum_{j}d_{j}}\frac{d_{i}}{\varepsilon
_{i}}U+g_{i}$.

Thus according to the above calculations, the general solution of
Eqs. (\ref {eq17}) is of the form

\begin{equation}
N_{i}^{+}=\frac{1}{2}\left ( {
\frac{Nc_{i}}{\sum_{j}c_{j}}+\frac{1}{\sum_{j}d_{j}}\frac{d_{i}}{\varepsilon
_{i}}U+f_{i}+g_{i}}\right ),\quad N_{i}^{-}=\frac{1}{2}\left
({\frac{Nc_{i}}{\sum_{j}c_{j}}-\frac{1}{\sum_{j}d_{j}}\frac{d_{i}}{\varepsilon
_{i}}U+f_{i}-g_{i}}\right ). \label{eq105}
\end{equation}
The parameters, $c_{i},$ $d_{i},$ $f_{i}$ and $g_{i}$, in the
general solution (\ref{eq105}) should be determined by using the
physical conditions of the true distribution. It is readily
verified that in the two-level system such as spin-$\frac{1}{2}$
model\cite{Datta,Mizrahi}, the parameters $c_{i},  d_{i}$ and
$f_{i}, g_{i}$ in the occupation numbers $N_{i}^{+}$ and
$N_{i}^{-}$ may be $c_{i}=d_{i}=\varepsilon _{i}$ and
$f_{i}=g_{i}=0$. Thus the form of $N_{i}^{+}$ and $N_{i}^{-}$ is
reduced to that of the familiar two-level system. This shows that
the above calculations are self-consistent.

According to the Boltzmann's entropy formula ( Boltzmann's
relation ), one has

\begin{equation}
S=k_{B}\ln \sum_{i}\frac{N!}{N_{i}^{+}!N_{i}^{-}}\simeq k_{B}
[N\ln N-\sum_{i}\left ({N_{i}^{+}\ln N_{i}^{+}+N_{i}^{-}\ln
N_{i}^{-}}\right ) ].
\end{equation}
By using  $\frac{\partial }{\partial U}\left ({N_{i}^{+}\ln
N_{i}^{+}}\right )=
\frac{1}{2}\frac{1}{\sum_{j}d_{j}}\frac{d_{i}}{\varepsilon
_{i}}\left ({\ln N_{i}^{+}+1}\right ), \quad \frac{\partial
}{\partial U}\left ({N_{i}^{-}\ln N_{i}^{-}}\right )=
-\frac{1}{2}\frac{1}{\sum_{j}d_{j}}\frac{d_{i}}{\varepsilon
_{i}}\left ({\ln N_{i}^{-}+1}\right )$, one can arrive at
\begin{eqnarray}
\frac{1}{T}=(\frac{\partial S}{\partial
U})_{D}&=&\frac{k_{B}}{2\sum_{j}d_{j}}\sum_{i}\frac{d_{i}}{\varepsilon
_{i}}\ln \frac{N_{i}^{-}}{N_{i}^{+}} \nonumber \\
&=&\frac{k_{B}}{2\sum_{j}d_{j}}\sum_{i}\frac{d_{i}}{\varepsilon
_{i}}\ln \left\{
\frac{\frac{Nc_{i}}{\sum_{j}c_{j}}-\frac{1}{\sum_{j}d_{j}}\frac{d_{i}}{\varepsilon
_{i}}[\sum_{l}(N_{l}^{+}-N_{l}^{-})\varepsilon
_{l}]+f_{i}-g_{i}}{\frac{Nc_{i}}{\sum_{j}c_{j}}+\frac{1}{\sum_{j}d_{j}}\frac{d_{i}}{\varepsilon
_{i}}[\sum_{l}(N_{l}^{+}-N_{l}^{-})\varepsilon
_{l}]+f_{i}+g_{i}}\right\}, \label{eq106}
\end{eqnarray}
where $D=\epsilon \epsilon _{0}E, \quad \varepsilon
_{i}=\frac{1}{\epsilon \epsilon _{0}}p_{0}D\cos \theta _{i}$. Here
in our tentative analysis we only consider the isotropic
homogeneous medium where the relation $D=\epsilon \epsilon _{0}E$
holds.

In the two-level system, both $c_{i}$ and $d_{i}$ are reduced to
$\varepsilon _{i}$. This, therefore, means that in general case,
$c_{i}$ and $d_{i}$ may also be the odd functions of $D$ ( or the
exterior electric field strength $E$ ). The reason lies in that
when the instantaneous inversion of the exterior electric field
$E$ takes place, some physical quantities will acquire a minus
sign, i. e., $E\rightarrow -E,c_{i}\rightarrow
-c_{i},d_{i}\rightarrow -d_{i}$, but $N_{i}^{+}$ and $N_{i}^{-}$
are unchanged. From the expressions (\ref {eq105}), it follows
that both $f_{i}+g_{i}$ and $f_{i}-g_{i}$ ( or both $f_{i}$ and
$g_{i}$ ) are the even functions with respect to $D$. By using
these results, it is shown that the absolute temperature $T$ in
the expression (\ref {eq106}) is the odd function with respect to
$D$. This result means that if $E\rightarrow -E$, then
$T\rightarrow -T$, namely, the instantaneous inversion of the
exterior electric field may cause the state of negative
temperature, which is analogous to the case in nuclear spin system
e. g., the nuclear spin system of the pure LiF crystal
\cite{Ramsey,Purcell,Klein}. It is therefore apparent that if the
swiftly varying exterior electric field inverses all of a sudden
and simultaneously the inversion of the dipole moments cannot
catch up with the rapid reversal of the exterior electric field,
then after it takes the relaxation time $\tau _{1}$ of the
dipole-dipole thermal equilibrium depending on the interaction
between the dipole moments in the exterior field, the
electric-dipole system is therefore getting into the state of
negative absolute temperature. If the relaxation time $\tau_{1}\ll
\tau_{2}$ with $\tau_{2}$ being the dipole-lattice relaxation time
of the thermal equilibrium between dipoles and lattices
\cite{Ramsey}, then this state of negative-temperature dipole
system can be readily realized in experiments. Note that this
effect via the instant reversals of electric fields can be
considered the first anti-shielding effect that acts on the
electric dipoles. Its essential feature is characterized by the
fact that the total potential energy of dipole system in the
reversed field is positive ( $U>0$ ). Here one may refer to the
case of $U<0$ as the normal shielding effect in electromagnetic
media.

It is verified that a second anti-shielding effect due to the
state of negative temperature acting upon the electromagnetic wave
occurs in the dielectric medium when the light is propagating
through it. In general the light field will acquire an effective
rest mass term resulting from the shielding effect of electric
dipoles. In the case of the instantaneously reversed fields, this
mass term in the field equation, however, possesses a minus sign
caused by the negative temperature $T$. The equation of the
electric field shielded by the electric dipoles may be given
$\nabla ^{2}\Phi =-\frac{Q}{\epsilon _{0}}\delta n$ with $Q$ being
the polarized effective charge in each electric dipole. The
right-handed side of this equation arises from the polarization
and shielding effect of electric dipoles and $\delta n=n[1-\exp
(\frac{Q\Phi }{k_{B}T})]$, where
$n=n_{0}\frac{p_{0}E}{k_{B}T}[\sinh (\frac{p_{0}E}{k_{B}T})]^{-1}$
with $n_{0}$ denoting the volume density of the electric dipoles
in the dielectric medium. We take the first-order approximation of
$\delta n$ and arrive at the result $\delta n\simeq -n\frac{Q\Phi
}{k_{B}T}$. Substitution of $\delta n$ into the above electric
field equation yields

\begin{equation}
\nabla ^{2}\Phi -\frac{nQ^{2}}{\epsilon _{0}k_{B}T}\Phi =0.
\label{eq21}
\end{equation}
It follows that the negative effective rest mass term,
$\frac{nQ^{2}}{\epsilon _{0}k_{B}T}\Phi $, in the field equation
(\ref {eq21}) of electric field arises from the anti-shielding
effect of electric dipoles due to the state of negative
temperature. Further calculation shows that the effective rest
mass of light field propagating inside this dielectric may be
expressed as

\begin{equation}
m_{eff}=\frac{\hbar }{c}\left ({\frac{nQ^{2}}{\epsilon
_{0}k_{B}T}}\right )^{\frac{1}{2}}.
                 \label{eq103}
\end{equation}
Hence we know that if the temperature, $T$, of the dipole system
becomes negative, then the effective rest mass ( that accordingly
becomes an imaginary number ) of light field may lead to the
superluminal light propagation in dielectric media. In the state
of low temperature, e. g., $T=1 {\rm K}$, if the instantaneous
inversion of the exterior electric field takes place, then the
order of magnitude of $\left| m_{eff}\right| $ may be
$10^{-31}{\rm kg }$, where without loss of generality $Q$ and $n$
are taken to be $10^{-20}{\rm C}$ and $10^{28}/{\rm m}^{3}$,
respectively. This means that once the typical light propagation
with $\hbar \omega =10^{-19}{\rm J}$ approaches the dielectric
medium in the state of negative temperature, the light field
obtains a negative rest mass term and therefore will exceed $c$.
The similar effect occurs when the supersymmetric partner of the
photon, i. e., photinos ( should such exist ) propagates in this
anti-shielding dielectric medium and will obtain a negative
effective rest mass, which results in the possible negative speed
( ${\bf {v}}<0$ ) against the propagation direction. It is of
interest that the negative speed is faster than the speed
exceeding $c$ ( even faster than the positively infinite velocity
), just similar to the fact that the state of negative temperature
is hotter than that of positively infinite temperature. In
addition, it should be noted that the potential superluminal light
propagation in the state of negative temperature ( or the
anti-shielding dielectric ) is not at odds with the special
relativity nor does it violate the principle of causality, since
the above theoretical analysis concerning the superluminal effect
is a direct consequence of both the Maxwell's theory and the
dynamics of the special relativity. Moreover, this superluminal
propagation with particle velocity exceeding $c$ is not strange to
us, since there also exists the `` tachyon '' representation with
the four-dimensional momentum squared $p^{\mu }p_{\mu }<0$ in the
Poincar$\acute{\rm e}$ \ group\cite{Ramond}, so that this
superluminal motion discussed above is natural from the
theoretical point of view. We hold that it deserved the
experimental investigation.

How to measure the velocity of this superluminal light propagation
in experiment? At low temperature, e. g., $T=1{\rm K},$ of the
dielectric such as the crystals of HCl and HF, which each dipole
respectively possesses $3.62\times 10^{-30}{\rm C\cdot m}$ and
$6.40\times 10^{-30}{\rm C\cdot m}$ of the dipole moment, in the
presence of exterior strong electric field that makes the electric
dipoles polarized, the instantaneous reversal of the exterior
field may cause the medium into the state of negative temperature.
In this situation one may let a light pulse produced at $t=0$
propagate a distance $l_{1}$ and then go through such polarized
medium with thickness $l_{2}$. If the light approaches the
detector at $t=t_{1}$, then one may take
$v=\frac{cl_{2}}{ct_{1}-l_{1\text{ }}}$ \cite{Zhou}, which exceeds
$c$. It is worthwhile to point out that in this superluminal
phenomenon the group velocity of light field made up of the
photons equals the particle velocity of the photons, since the
effective rest mass $m_{eff}=\frac{\hbar }{c}\left
({\frac{nQ^{2}}{\epsilon _{0}k_{B}T}}\right )^{\frac{1}{2}}$ is
independent of the frequency $\omega $ of light field. This,
therefore, implies that the group velocity of the light pulse
propagating in the anti-shielding dielectric medium may also
exceed $c$ .

To close this section we briefly consider the anti-shielding
effect of thermal electron plasma. According to the Debye's
shielding effect\cite{Cairns}, the Debye's length is expressed by
$\lambda _{D}=\left( \frac{\epsilon _{0}k_{B}T}{N(e^{\ast
})^{2}}\right) ^{\frac{1}{2}}$ with $N$ being the electron number
in per unit volume; hence, the effective rest mass of light field
due to this shielding effect is $m_{1}=\frac{\hbar }{c}\left(
\frac{N(e^{\ast })^{2}}{\epsilon _{0}k_{B}T}\right)
^{\frac{1}{2}}$. If, for example, the plasma is placed in an
exterior electric field, here the temperature $T$ in the
expression for $m_{1}$ is that of the thermal equilibrium caused
by potential-potential interaction between electrons themselves (
and ions ) in the electric field, where the particle distribution
is in the real space ( associated with the electric potential
$\phi \left( {\bf x}\right) $ given by exterior field ), rather
than in the velocity space ( associated with the Maxwell's
distribution law ).  As is presented previously, in cold plasma
the light field acquires an effective rest mass whose square is
$m_{2}^{2}=\frac{\hbar ^{2}}{c^{2}}\frac{N(e^{\ast
})^{2}}{\epsilon _{0}}\frac{1}{mc^{2}}$. This result can also
apply to the thermal plasma. It thus follows that in thermal
electron plasma the total square of rest mass of light field is
written in the form

\begin{equation}
m_{total}^{2}=m_{1}^{2}+m_{2}^{2}=\frac{\hbar
^{2}}{c^{2}}\frac{N(e^{\ast })^{2}}{\epsilon _{0}}\left(
\frac{1}{k_{B}T}+\frac{1}{mc^{2}}\right) ,\label{eq39}
\end{equation}
where $T$ is the temperature of electrons in potential-potential
thermal equilibrium in the real space ( configuration space ). In
general this temperature differs from the Maxwell's temperature
that depends upon the distribution of particle velocity ( in
velocity space ). This thermal equilibrium ( Maxwell's velocity
distribution ) is often attained by, for instance, the
electron-ion collisions where the relaxation time is proportional
to $N^{-1}$. This relaxation process to thermal equilibrium can be
described by a kinetic equation\cite{Alexandrov}. Once if the
potential-potential relaxation time ( proportional to
$N^{-\frac{1}{3}}$ ) is far less than the electron-ion relaxation
time, then in the process of the instantaneous reversal of
exterior electric field, the potential-potential system is
therefore thermally isolated from the Maxwell's velocity
distribution system. It follows from (\ref{eq39}) that if the
negative temperature $T$ caused by the instantaneous reversal of
exterior field satisfies $0^{-}>T>-\frac{mc^{2}}{k_{B}}$, then the
total effective mass squared $m_{total}^{2}<0$. However, it should
be noted that in the case of thermal plasma the negative
temperature is not easily achieved, since the potential-potential
system consisting of both quasi-free electrons and ions possesses
excessive free energy due to the stored positive potential energy
of charged particles in reversed electric field and is therefore
not very stable\cite{Alexandrov}. Only a rather weak perturbation
would necessarily destroy this negative-temperature system. For
this reason we are therefore not interested in further considering
the anti-shielding effect of thermal plasma and the related
phenomena.

\section{Anti-shielding effect and negative temperature in left-handed media}
More recently, a kind of composite media ( the so-called
left-handed media ) having a frequency band where the {\it
effective permittivity} ( $\varepsilon$ ) and the {\it effective
permeability} ( $\mu$ ) are simultaneously negative attracts
attentions of many researchers in various fields such as materials
science, condensed matter physics, optics and
electromagnetism\cite{Smith,Klimov,Pendry3,Shelby,Ziolkowski2}.
Veselago first considered this peculiar medium and showed that it
possesses a negative index of refraction\cite{Veselago}. It
follows from the Maxwell's equations that in this medium the
Poynting vector and wave vector of electromagnetic wave would be
antiparallel, i. e., the vector {\bf {k}}, the electric field {\bf
{E}} and the magnetic field {\bf {H}} form a left-handed system;
thus Veselago referred to such materials as ``left-handed'', and
correspondingly, the ordinary medium in which {\bf {k}}, {\bf {E}}
and {\bf {H}} form a right-handed system may be termed the
``right-handed'' medium. Other authors call this class of
materials ``negative-index media (NIM)'', ``double negative media
(DNM) \cite{Ziolkowski2}'' and Veselago's media. It is readily
verified that in such media having both $\varepsilon$ and $\mu$
negative, there exist a number of peculiar electromagnetic
properties, for instance, many dramatically different propagation
characteristics stem from the sign change of the group velocity,
including reversal of both the Doppler shift and Cherenkov
radiation, anomalous refraction, modified spontaneous emission
rates and even reversals of radiation pressure to radiation
tension\cite{Klimov}. In experiments, this artificial negative
electric permittivity media may be obtained by using the array of
long metallic wires (ALMWs), which simulates the plasma behavior
at microwave frequencies, and the artificial negative magnetic
permeability media may be built up by using small resonant
metallic particles, e. g., the split ring resonators (SRRs), with
very high magnetic
polarizability\cite{Pendry1,Pendry2,Pendry3,Maslovski}.

In linear media the electric and magnetic polarizations ( per unit
volume ) is respectively

\begin{equation}
{\bf P}=(\epsilon -1)\epsilon _{0}{\bf E}, \quad {\bf M}=(\mu
-1){\bf H}.
\end{equation}
The potential energy density of electric and magnetic dipoles in
electromagnetic fields reads

\begin{equation}
U_{e}=-{\bf P}\cdot {\bf E}=-(\epsilon -1)\epsilon _{0}{\bf
E}^{2},\quad U_{m}=-\mu _{0}{\bf M}\cdot {\bf H}=-(\mu -1)\mu
_{0}{\bf H}^{2}   \label{eqq51}
\end{equation}
with $\mu _{0}$ being the permeability of free space. It is
apparent that in left-handed media where both $\epsilon $ and $\mu
$ are negative, the electric and magnetic polarizations, ${\bf P}$
and ${\bf M}$, are opposite to the electric and magnetic fields,
${\bf E}$ and ${\bf H}$, thus in general, the potential energy
density, both $U_{e}$ and $U_{m}$, are positive, which differs
from the cases in the conventional media, i. e., the right-handed
media. This, therefore, means that left-handed media possess the
anti-shielding effect, which results in many anomalous
electromagnetic and optical behaviors\cite{Klimov}, as is
mentioned above.

May the electric and magnetic dipole systems be brought into the
state of negative absolute temperature when the electromagnetic
wave is propagating in left-handed media? We show that under some
restricted conditions, these dipole systems may possess a negative
temperature indeed. The detailed analysis is given in what
follows:

According to Ramsey's claim\cite{Ramsey} that the essential
requirement for a thermodynamical system to be capable of a
negative temperature is that the elements of the thermodynamical
system must be in thermodynamical equilibrium among themselves in
order that the system can be described by a temperature at all.
For the left-handed media, this may be understood in the sense:
the electric and magnetic dipole moments produced by ALMWs and
SRRs should be coupled respectively to each other, and this
dipole-dipole interactions must be strong and therefore the
thermodynamical equilibriums between themselves may be brought
about in a short time, which is characterized by the relaxation
time $\tau _{1}$. But here both ALMWs and SRRs are macroscopic
elements of left-handed medium ( e. g., the length scale of SRRs
is mm ), which implies that there exists a problem as to whether
the concept of thermodynamical equilibrium is applicable to these
macroscopic elements or not. This problem, however, is not the
subject of this section, and it will be published in a separate
paper; furthermore, if, for example, we can find a kind of
molecular-type SRRs ( i. e., the C- or $\rm {\Omega}$-
like\cite{Saadoun} curved organic molecule that possesses the
delocalized $\pi $ bond and can therefore provide a molecular
electric current induced by magnetic field of lightwave ) for
manufacturing the left-handed media, then the dipole system of
this media can be appropriately described by a
temperature\cite{Ramsey}. So in our tentative consideration in
this paper, we can study the anti-shielding effect and negative
temperature of left-handed media from the phenomenological point
of view, ignoring the detailed information about the elemental
atoms which constitute the left-handed media.

In order to generate a state of negative absolute temperature, the
following requirement for a thermodynamical system should also be
satisfied: the system must be thermally isolated from all other
systems, namely, the thermal equilibrium time ($\tau _{1}$) among
the elements of the negative-temperature system must be short
compared to the time during which appreciable energy is lost to or
gained from other systems\cite{Ramsey}. In left-handed media, the
dipole-lattice interaction causes heat to flow from the dipole
system ( negative temperature ) to the lattice system ( positive
temperature ). If the relaxation time ($\tau _{2}$) during which
this interaction brings the dipole-lattice system into
thermodynamical equilibrium is much greater than $\tau _{1},$ then
it may be said that the dipole system is thermally isolated from
the lattice system, and the dipole system is therefore capable of
a negative temperature. In addition to this condition, there
exists another restriction, i. e., the state of negative
temperature should be built up within half period of
electromagnetic wave propagating inside the left-handed medium, i.
e., the relaxation time $\tau _{1}$ should be much less than
$\frac{1}{f}$ with $f$ being the frequency of  electromagnetic
wave, which means that this thermodynamical process caused by
dipole-wave interaction can be considered a quasi-stationary
process and in consequence the state of negative temperature of
dipole system due to dipole-dipole interaction may adiabatically
achieve establishment in every instantaneous interval during one
period of electromagnetic wave. If this condition is not
satisfied, say, the relaxation time $\tau _{1}$ is several times
as long as $\frac{1}{f}$, then the negative temperature of dipole
systems cannot be achieved since the thermodynamical equilibrium
among dipoles induced by the oscillating electromagnetic fields
${\bf E}$ and ${\bf H}$ may lose its possibility, namely, the
concept of thermodynamical equilibrium is not applicable to this
non-stationary process; in fact, it should be treated by using
non-equilibrium statistical mechanics. In view of above
discussions, we conclude that the principal conditions for the
electric and magnetic dipole systems of left-handed media
possessing a negative temperature may be ascribed to the following
two essential requirements

\begin{equation}
\tau _{1}\ll \tau _{2},\quad \tau _{1}\ll \frac{1}{f}.
\end{equation}

Generally speaking, the first restriction, $\tau _{1}\ll \tau
_{2}$, is readily achieved, since in electromagnetic media the
dipole-dipole interaction are often much stronger than the
dipole-lattice interaction. For example, in some nuclear spin
systems such as the pure crystal of LiF where the negative
temperature was first realized in experiments, the relaxation time
$\tau _{1}$ of spin-spin process is approximately the period of
the Larmor precession of one nucleus in the field of its neighbor
and is of the order of $10^{-5}$ second while the relaxation time
that depends strongly upon the interaction between the spin system
and the crystal lattice is several minutes\cite{Ramsey}. In the
recent work of designing the structures of more effective
left-handed media, use is made of the SRRs of high electromagnetic
polarizability\cite{Smith}. This leads to the very strong
dipole-dipole interaction and may achieve the thermodynamical
equilibrium of dipole system with shorter relaxation time. If this
interaction is rather strong and therefore agrees with the second
restriction, $\tau _{1}\ll \frac{1}{f}$ with $f$ being in the
region of GHz\cite{Pendry1,Pendry2,Pendry3}, then the dipole
systems of left-handed media may be said to be in the state of
negative absolute temperature.

Since the most conspicuous feature of left-handed media may be the
electromagnetic anti-shielding effect, the previous considerations
concerning the superluminal light propagation in the instantly
varying electric fields based on the viewpoint of effective rest
mass can also be applied to the incident electromagnetic wave
propagating inside left-handed media. As is stated, in left-handed
media many electromagnetic effects such as Doppler shift,
Cherenkov radiation and wave refraction are inverted compared with
those in ordinary right-handed media. In the similar fashion, it
is believed that this superluminal light propagation may also
probably occur in left-handed media. Moreover, in some
electromagnetic materials, there exists another superluminal pulse
propagation that also results from the anti-shielding
effect\cite{Ziolkowski3,Ziolkowski4} ( but rather than from the
statistical fluctuations in negative temperature ), since in this
media the designed equivalent permittivity and permeability based
on the so-called two-time-derivative Lorentz model are
simultaneously smaller than the values in free
space\cite{Ziolkowski3,Ziolkowski4}. Ziolkowski has studied the
superluminal pulse propagation and consequent superluminal
information exchange in this media, and demonstrated that they do
not violate the principle of causality\cite {Ziolkowski}. We
suggest that all these potential superluminality effects of light
propagation in anti-shielding media should be further
investigated, particularly in experiments.

\section{Concluding remarks}
In electromagnetic systems, the anti-shielding effect is not
easily found. The normal shielding effect makes the speed of light
wave propagating inside media less than that in a vacuum. In these
media the directions of induced electric and magnetic dipoles are
parallel ( not antiparallel ) to the electromagnetic fields, and
hence the potential energy of these dipoles in the fields are
negative as is stated in (\ref{eqq51}). It is for these reasons
that in electromagnetic media, systems of negative temperature
occur only rarely. In the present paper, for the case of the
instantaneous reversal of exterior electric field, we take into
consideration the state of negative temperature and the
anti-shielding effect of electric dipoles. It follows that if the
energy $\varepsilon _{i}$ is changed into $-\varepsilon _{i}$, i.
e., once the external field strength changes the sign, after
taking the relaxation time the temperature might accordingly
become negative ( additionally, in the meanwhile the restriction
of the dipole-dipole relaxation time being much shorter than the
dipole-lattice relaxation time should also be satisfied ).
Meanwhile, it follows from the expression for the effective rest
mass $m_{eff}=\frac{\hbar }{c}\left ({\frac{nQ^{2}}{\epsilon
_{0}k_{B}T}}\right )^{\frac{1}{2}}$ that if $T\rightarrow -T$,
then $m_{eff}$ may become imaginary. Hence it is shown from the
point of view of effective rest mass that the potentially
interesting superluminal light propagation in media may result
from this anti-shielding effect of dielectric in the state of
negative temperature. According to the conventional viewpoints,
the `` tachyon '' representation in the Poincar$\acute{\rm e}$
group is often considered the non-physical representation. On the
contrary, however, we favor that probably in some certain
anti-shielding cases ( including the negative-temperature state ),
this potential superluminal propagation of light would exist,
which deserves the further investigation.

For the case of left-handed media, the anti-shielding effect and
the requirements for left-handed media to be capable of negative
temperature is analyzed in this paper. In recent investigations
regarding the manufacture of left-handed materials, the media
capable of negative index of refraction is restricted to the
microwave frequency region, since the resonance frequency of SRRs
( and ALMWs ) is low (GHz). If one can find the molecular-type
SRRs by certain artificial molecule arrangements, which has the
conjugation effect of organic molecules due to the delocalized
$\pi $ bond and the consequent high magnetic susceptibility, then
the resonance frequency would be readily raised greatly by several
orders of magnitude. Based on this, it is of physical significance
to consider the interaction of radiation fields with
negative-temperature systems in left-handed media. Klimov
theoretically studied the processes of spontaneous emission of an
atom placed near a body ( sphere ) made of the Left-handed
material and placed in the right-handed medium, and showed that
these processes drastically differ from those near usual
right-handed body\cite{Klimov}. We think if the statistical effect
( including negative temperature ) of left-handed is taken into
account, then the results obtained by Klimov should be modified.
This modification is under consideration based on the combination
of Klimov's formulation and the thermodynamics and statistical
mechanics.

Acknowledgements The author thanks Q. Wu, T. Chang and X. M. Wu
for their beneficial discussions concerning the superluminal light
propagation. I have benefited from useful criticism and advice
from C. Alden Mead. My special thanks are due to him also. This
project is supported in part by the National Natural Science
Foundation of China under the project No. $90101024$.

\end{document}